\documentclass[twocolumn,twoside,slac_two]{revtex4}
\usepackage{graphicx}
\usepackage{fancyhdr}
\begin{document}

\title{SWIFT and BATSE bursts' classification}

%

\author{I. Horv\'ath}
\affiliation{Bolyai Military University, Budapest, Hungary }
\author{Z. Bagoly}
\affiliation{E\"otv\"os University, Budapest, Hungary }

\author{L.G. Bal\'azs}
\affiliation{Konkoly Observatory, Budapest, Hungary }

\author{G. Tusn\'ady}
\affiliation{R\'enyi Institute of Mathematics, Budapest, Hungary  }

\author{P. Veres}
\affiliation{Bolyai Military University and E\"otv\"os University, Budapest, Hungary }

\begin{abstract}
Two classes of gamma-ray bursts were identified in 
the BATSE catalogs characterized by their durations. 
There were also some indications for the existence 
of a third type of gamma-ray bursts. Swift satellite 
detectors have different spectral sensitivity than 
pre-Swift ones for GRBs. Therefore in  this paper we 
analyze the bursts' duration distribution and also the 
duration-hardness bivariate distribution, published in 
The First BAT Catalog. Similarly to the BATSE data, 
to explain the BAT GRBs' duration distribution three 
components are needed. Although, the relative 
frequencies of the groups are different than 
they were in the BATSE GRB sample, the difference 
in the instrument spectral sensitivities can 
explain this bias in a natural way. 
This means theoretical models may have 
to explain three different type of gamma-ray bursts. 

\end{abstract}

\maketitle

\thispagestyle{fancy}


\begin{figure*}[t]
\centering
\includegraphics[angle=270,width=135mm]{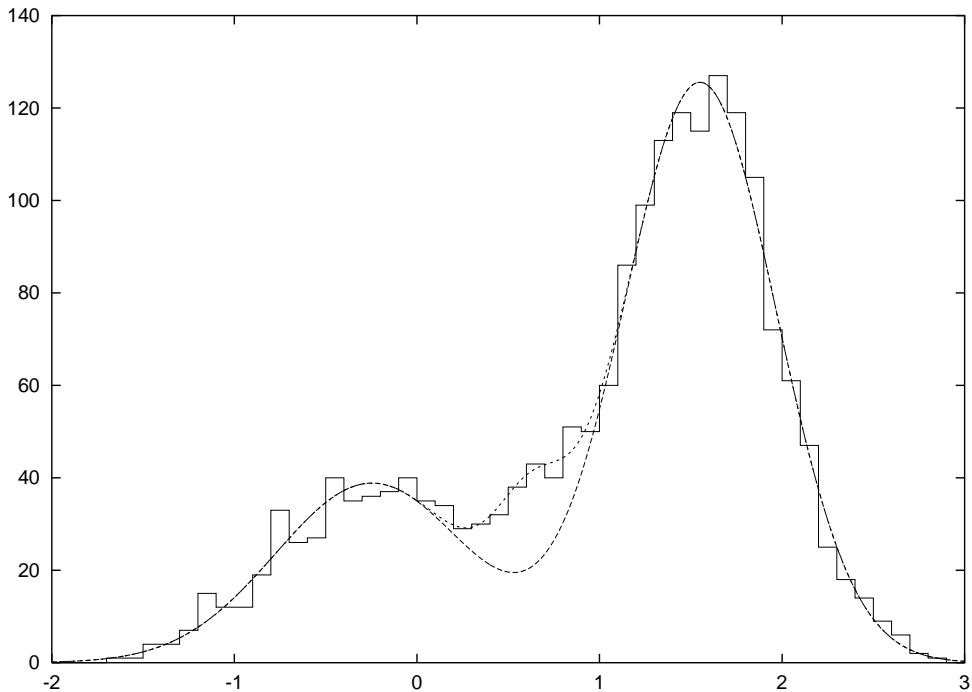}
\caption{The duration distribution of the BATSE bursts and the two- and three-Gaussian fits.} \label{f1}
\end{figure*}

\bigskip

\bigskip

\bigskip

\bigskip

\bigskip

\bigskip

\bigskip

\section{INTRODUCTION}

The discovery of the third type of GRBs goes back as early
as 1998 \cite{muk98,hor98}. After that many research groups studied
the BATSE bursts' sample and concluded the third
group of the GRBs statistically exists
\cite{hak00,bala01,rm02,hor02,hak03,bor04,hor06,chat07}.
Later several papers were published analysing
different data sets \cite{hor08,rmw09}.
In this paper we 
analyze the bursts' duration distribution and also the 
duration-hardness bivariate distribution, published in 
The First BAT and The BATSE Catalog.

\section{THE ONE DIMENSIONAL GAUSSIAN FITS}

We used the Maximum Likelihood (ML) method for the
analysis of the (Swift) BAT and BATSE bursts.
The ML method assumes that the probability density function of an
$x$ observable variable is given in the form of $g(x,p_1,...,p_k)$
where $p_1,...,p_k$ are parameters of unknown value. Having  $ N $
observations on $x$ one can define the likelihood function in the
following form:

\begin{equation}
l = \prod_{i=1}^{N}  g(x_i,p_1,...,p_k) \
\end{equation}

\noindent or in logarithmic form (the logarithmic form is more
convenient for calculations):

\begin{equation}
L= \log l = \sum_{i=1}^{N} \log  \left( g (x_i,p_1,...,p_k) \right)
\end{equation}

The ML procedure maximizes $L$ according to
$p_1,...,p_k$.  Since the logarithmic function is monotonic the
logarithm reaches the maximum where $l$ does it as well.
The confidence region of the estimated parameters is given by the
following formula, where $L_{max}$ is the maximum value of the
likelihood function and $L_0$ is the
likelihood function  at the true value of the
parameters \citep{KS73}:

\begin{equation}
2 ( L_{max} -  L_0) \approx \chi^2_k \label{eq:chi}
\end{equation}

Therefore one can fit the $\log T_{90}$  distribution using
ML with a superposition of $k$ log-normal components, each
of them having 3 unknown parameters to be fitted with $N$
measured points. Our goal is to find the minimum value
of $k$ suitable to fit the observed distribution. Assuming a
weighted  superposition of $k$ log-normal distributions one has to
maximize the following likelihood function:

\begin{equation}
L_k = \sum_{i=1}^{N} \log  \left(\sum_{l=1}^k   w_lf_l(x_i,\log
T_l,\sigma_l ) \right)
\end{equation}

\noindent where $w_l$ is a weight, $f_l$ a log-normal function with
$\log T_l$ mean and $\sigma_l $ standard deviation having the form
of

\begin{equation}
f_l = \frac{1}{ \sigma_l  \sqrt{2 \pi  }}
\exp\left( - \frac{(x-\log T_l)^2}{2\sigma_l^2} \right)  
\label{fk}
\end{equation}

\noindent and due to a normalization condition

\begin{equation}\label{wight}
    \sum_{l=1}^k w_l= N \, 
\end{equation}

\bigskip

\section{ONE DIMENSIONAL GAUSSIAN FITS FOR THE BATSE BURSTS}

A fit to the duration distribution of the BATSE bursts was taken using a maximum likelihood method  with the superposition of two  log-normal distributions. 
This can be done by a standard search for  5 parameters with $N=1929$ measured points.  
Both log-normal distributions have two parameters; the fifth parameter  defines the weight ($w_1$) of the first log-normal distribution. The second weight is $w_2=(N-w_1)$ due to the normalization. 
Therefore we obtain the best fit to the 5 parameters through a maximum likelihood  estimation.
 The maximum of the likelihood was $L_2$=12320.11.  
 
Secondly, a three-Gaussian fit was made (Figure 1. shows both the two and the three component fits)
with three $f_k$ functions with eight parameters (three means, three standard deviations and two weights).
The logarithm of the best  likelihood ($L_3$) is 12326.25.     
According to the mathematical theory, twice the difference of these  numbers follows the $\chi  ^2$ distribution with three degrees of freedom because the new fit has three more parameters.
The difference is 6.14 which gives us a   0.5\% probability.
Therefore the three-Gaussian fit is better and  there is only
a 0.005 probability that it is caused by statistical fluctuation.

\begin{figure*}[t]
\centering
\includegraphics[width=135mm]{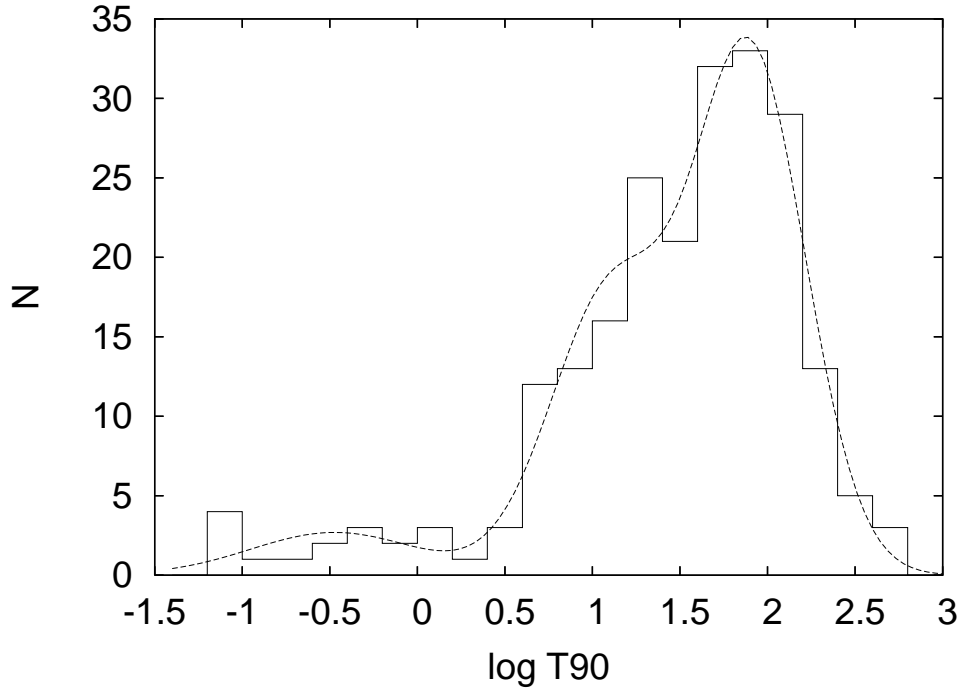}
\caption{The duration distribution of the Swift bursts and a three-Gaussian fit.} \label{f2}
\end{figure*}

\bigskip

\section{ONE DIMENSIONAL GAUSSIAN FITS FOR SWIFT GRBS}

In the Swift BAT Catalog \citep{sak08} there are 237 GRBs, of
which 222 have duration information. Fig. 2.  shows the
$\log T_{90}$ distribution. We made fits for this distribution.
 Assuming only one log-normal
component the fit gives $L_{1max}=951.666$ but in the case of
$k$=2 one gets $L_{2max}=983.317$ .

Based on Eq. (\ref{eq:chi}) we can infer whether the addition of a
further log-normal component is necessary to signifincantly improve
the fit. We make the null hypothesis that we have reached already
the the true value of $k$. Adding a new component, i.e. moving
from $k$ to $k+1$, the ML solution of $L_{kmax}$ has changed to
$L_{(k+1)max}$, but $L_0$ remained the same. In the meantime we
increased the number of parameters with 3 ($w_{k+1}$, $logT_{k+1}$
and $\sigma_{(k+1)})$. Applying Eq. (\ref{eq:chi}) on both
$L_{kmax}$ and $L_{(k+1)max}$ we get after subtraction

\begin{equation}\label{kk1}
2 ( L_{(k+1)max} -  L_{kmax}) \approx \chi^2_3 \, 
\end{equation}

\noindent For $k=1$ $L_{2max}$ is greater than $L_{1max}$
by more than 30, which gives for $\chi^2_3$ an
extremely low probability of $5.88\times 10^{-13}$. It means the
two log-normal fit is really a better approximation for
the duration distribution of GRBs than one log-normal.

Thirdly, a
three-log-normal fit was made combining three $f_k$ functions with
eight parameters (three means, three standard deviations and two
weights).   The highest value of the
logarithm of the likelihood ($L_{3max}$) is 989.822.  For two
log-normal functions the maximum  was $L_{2max}=983.317$.  The maximum thus
improved by 6.505. Twice of this is 13.01 which gives us the
probability of 0.46\% for the difference between $L_{2max}$
and $L_{3max}$ is being only by chance. Therefore there is only a
small chance the third log-normal is not needed. Differently said the
three-log-normal fit (see Figure 2.) is better and there is a 0.0046
probability that it was caused only by statistical fluctuation.

One should also calculate the likelihood for
four log-normal functions. The best logarithm of the ML is 990.323.
It is bigger with 0.501 than it was with three log-normal functions.
This gives us a low significance (80.1\%),
therefore the fourth component is not needed. 

\bigskip

\bigskip

\bigskip

\section{THE TWO DIMENSIONAL GAUSSIAN FITS}

\begin{figure*}[t]
\centering
\includegraphics[angle=270,width=135mm]{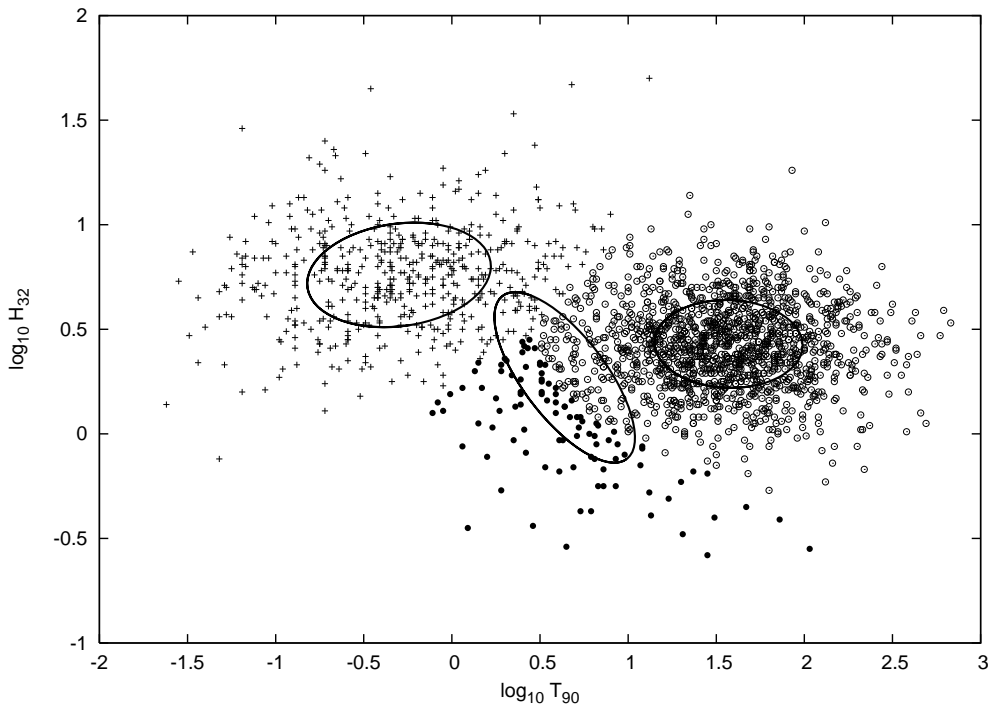}
\caption{The duration - hardness distribution of the BATSE bursts and the three components.} \label{f3}
\end{figure*}

When studying a GRB distribution, one can assume that the
observed probability distribution in the  parameter space is a
superposition of the distributions characterizing the different
types of bursts present in the sample. Using the notations $x$
and $y$ for the variables (in a 2D space), and using the law of
full probabilities, one can write
\begin{equation}\label{lfpr}
    p(x,y)
    =\sum \limits_{l=1}^k p(x,y|l)p_l
\end{equation}
In this equation $p(x,y|l)$ is the conditional probability
density assuming that a burst belongs to the $l$-th class. $p_l$
is the probability for this class in the observed sample ($\sum
\limits_{l=1}^k p_l = 1$), where $k$ is the number of classes.
In order to decompose the observed probability distribution
$p(x,y)$ into the superposition of different classes we need the
functional form of $p(x,y|l)$. The probability distribution of
the logarithm of durations can be well fitted by Gaussian
distributions, if we restrict ourselves to the short and long
GRBs \citep{hor98}. We assume the same also for the $y$
coordinate. With this assumption we obtain, for a certain $l$-th
class of GRBs,

\begin{widetext}

\begin{equation} \label{gauss}
p(x,y|l)  =
 \frac{1}{2 \pi \sigma_x \sigma_y
\sqrt{1-r^2}} \times 
\exp\left[-\frac{1}{2(1-r^2)}
\left(\frac{(x-a_x)^2}{\sigma_x^2} + \frac{(y-a_y)^2}{\sigma_y^2}
- \frac{2r(x-a_x)(y-a_y)} {\sigma_x \sigma_y}\right)\right], \;
\end{equation}
\end{widetext}

where  $a_x$, $a_y$ are the means, $\sigma_x$, $\sigma_y$ are
the dispersions, and $r$ is the correlation coefficient.
  Hence, a certain class is defined by 5 independent
parameters, $a_x$, $a_y$, $\sigma_x$, $\sigma_y$, $r$, which are
different for different $l$. If we have $k$ classes, then we
have $(6k - 1)$ independent parameters (constants), because any
class is given by the five parameters of Eq.(\ref{gauss}) and
the weight $p_l$ of the class. One weight is not independent,
because  $\sum \limits_{l=1}^{k} p_l = 1$.  The sum of $k$
functions defined by Eq.(\ref{gauss}) gives the theoretical
function of the fit.

By decomposing $p(x,y)$ into the superposition of $p(x,y|l)$
conditional probabilities one divides the original population of
GRBs into $k$ groups.
Decomposing the left-hand side of Eq.(\ref{lfpr}) into the sum
of the right-hand side, one needs the functional form of
$p(x,y|l)$ distributions, and also $k$ has to be fixed. Because
we assume that the functional form is a bivariate Gaussian
distribution (see Eq.(\ref{gauss})), our task is reduced to
evaluating its parameters, $k$ and $p_l$.

\citet{bal03} used this method for $k=2$, and gave a more
detailed description of the procedure. 
  However, that paper used fluence instead of
hardness, and used the BATSE data.  Here we will make similar
calculations for $k=2$, $k=3$ and $k=4$ using CGRO and Swift
observations.

\section{THE TWO DIMENSIONAL GAUSSIAN FITS FOR BATSE BURSTS}

The  data from The Final BATSE GRB Catalog
have been used, in which there are 2702 GRBs, for 1956
of which both the  hardnesses and durations are measured. These
1956 GRBs define the sample studied in here.
The  hardness duration distribution can be seen in Figure 3.

\begin{figure*}[t]
\centering
\includegraphics[width=135mm]{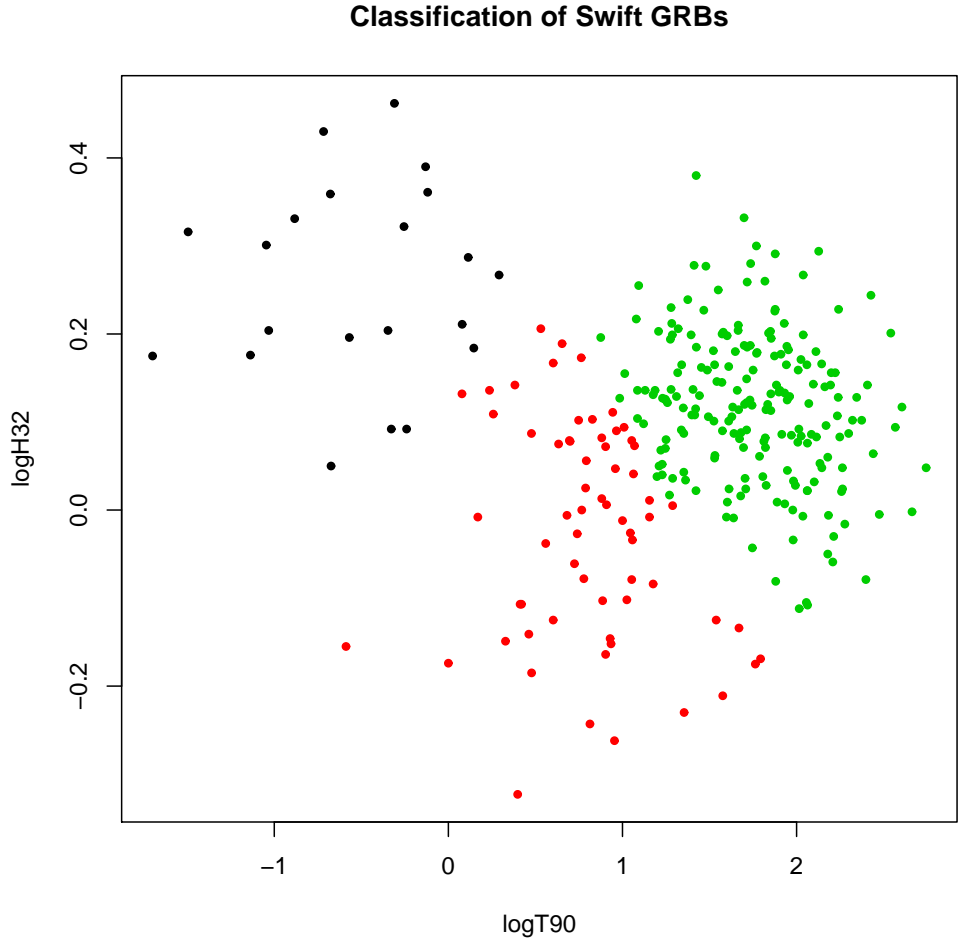}
\caption{The duration - hardness distribution of the Swift bursts and the three components.} \label{f4}
\end{figure*}

Moving from $k=2$ to $k=3$ the number of
parameters $m$  increases by 6
 (from 11 to 17), and $L_{max}$  grows from
1193 to 1237. Since $\chi_{17}^2=\chi_{11}^2+\chi_6^2$ the
increase in $L_{max}$ by a value of 44 corresponds to a value of
88 for a $\chi_6^2$ distribution. The probability for $\chi_6^2
\geq 88$ is extremely low ($<10^{-10}$). Therefore we may conclude that
the inclusion of a third class into the fitting procedure is well
justified by a very high level of significance.

Moving from $k=3$ to $k=4$, however, the improvement in $L_{max}$
is only 6 (from 1137 to 1143) corresponding to $\chi_6^2 \geq 12$,
which can happen by chance with a probability of 6.2 \%. Hence,
the inclusion of the fourth class is {\it not} justified. We may
conclude from this analysis that the superposition of three
Gaussian bivariate distributions - and {\it only these three ones} -
can describe the observed distribution of the BATSE data
Figure 3. also shows the three components.

\section{THE TWO DIMENSIONAL GAUSSIAN FITS FOR THE SWIFT BURSTS}

In the Swift BAT Catalog \citep{sak08} there are 237 GRBs, of
which 222 have duration information. Following the 
same procedure of data reduction we have extended this
sample with all the bursts detected until mid December 2008
(ending with GRB 081211a).  
Our total sample thus comprises the first four years of the
Swift satellite (since the detection of its first burst GRB
041217) and includes $342$ bursts.  $222$ from \citet{sak08} and
$120$ reduced by us.  The data reduction was done by using
HEAsoft v.6.3.2 and calibration database v.20070924. For
lightcurves and spectra we ran the \texttt{batgrbproduct}
pipeline.  We fitted the spectra integrated for
the duration of the burst with a power law model and a power law
model with an exponential cutoff. As in \citet{sak08} we have
chosen the cutoff power law model if the $\chi^2 $ of the fit
improved by more than 6.

For calculating the hardness ratio we have chosen fluence 2
($25-50 keV$) and fluence 3 ($50-100 keV$) and the hardness is
defined by the $HR = F3 / F2 $ ratio.
The duration - hardness distribution can be seen in Figure 4.
The two dimensional fits have been made in this plane.

Moving from $k=2$ to $k=3$ the number of
parameters $m$ increases by 6 (from 11 to 17), and $L_{max}$
grows from 506.6 to 531.4.  
Since $\chi_{17}^2=\chi_{11}^2+\chi_6^2$ the increase in
$L_{max}$ by a value of 25 corresponds to a value of 50 for a
$\chi_6^2$ distribution.  The probability for $\chi_6^2 \geq 50$
is very low ($10^{-8}$), therefore we  conclude that the inclusion
of a third class into the fitting procedure is well justified by
a very high level of significance.

Moving from $k=3$ to $k=4$, however, the improvement in
$L_{max}$ is 3.4 (from 531.4 to 534.8) corresponding to
$\chi_6^2 \geq 6.8$, which can happen by chance with a
probability of 33.9 \%. 
Hence, the inclusion of the fourth class is {\em not} justified.
We conclude again that the superposition of
three Gaussian bivariate distributions - and {\em only these
three ones} - can describe the observed distribution
of the Swift BAT bursts.

One can see from the fits and olso from Figure 4.
that the mean hardness of the intermediate class is very low -
the third class is the softest one. This is  in a good agreement
with the BATSE fits (for more details see \citet{hor06}), where we found that the intermediate duration class
is the softest in the BATSE database.  In that database 
11\% of all GRBs belonged to this group. 
In our analysis $p_2 =0.296$, therefore 30\% 
of the Swift bursts belong to the third 
(intermediate duration)group.

\begin{acknowledgments}
This research is supported by Hungarian OTKA grant K077795, 
and  by a Bolyai Scholarship (I.H.).  
\end{acknowledgments}

	    \bibliography{horv}

\end{document}